# Virtual Planetary Space Weather Services offered by the Europlanet H2020 Research Infrastructure

N. André[1], M. Grande[2], N. Achilleos[3], M. Barthélémy[4], M. Bouchemit[1], K. Benson[3], P.-L. Blelly[1], E. Budnik[1], S. Caussarieu[5], B. Cecconi[6], T. Cook[2], V. Génot[1], P. Guio[3], A. Goutenoir[1], B. Grison[7], R. Hueso[8], M. Indurain[1], G. H. Jones[9,10], J. Lilensten[4], A. Marchaudon[1], D. Matthiäe[11], A. Opitz[12], A. Rouillard[1], I. Stanislawska[13], J. Soucek[7], C. Tao[14], L. Tomasik[13], J. Vaubaillon[6]

[1]Institut de Recherche en Astrophysique et Planétologie, CNRS, Université Paul Sabatier, Toulouse, France (nicolas.andre@irap.omp.eu)
[2]Department of Physics, Aberystwyth University, Wales, UK
[3]University College London, London, UK
[4]Institut de Planétologie et d'Astrophysique de Grenoble, UGA/CNRS-INSU, Grenoble, France
[5]GFI Informatique, Toulouse, France
[6]LESIA, Observatoire de Paris, CNRS, UPMC, University Paris Diderot, Meudon, France
[7]Institute of Atmospheric Physics (IAP), Czech Academy of Science, Prague, Czech Republic
[8]Departamento de Física Aplicada I, Escuela de Ingeniería de Bilbao, Universidad del País Vasco UPV /EHU, Bilbao, Spain
[9]Mullard Space Science Laboratory, University College London, Holmbury Saint Mary, UK
[10]The Centre for Planetary Sciences at UCL/Birkbeck, London, UK
[11]German Aerospace Center (DLR), Institute of Aerospace Medicine, Linder Höhe, 51147 Cologne, Germany
[12]Wigner Research Centre for Physics, Budapest, Hungary
[13]Space Research Centre, Polish Academy of Sciences, Warsaw, Poland
[14]National Institute of Information and Communications Technology (NICT) 4-2-1, Nukui-Kitamachi, Koganei, Tokyo 184-8795, Japan

**Highlights:**
- Planetary Space Weather Services are currently developed within the Europlanet H2020 Research Infrastructure
- Planetary Space Weather Services will use a few tools and standards developed for the Astronomy Virtual Observatory
- We illustrate several VO-compliant functionalities already implemented in various tools

**Keywords:**
Virtual Observatory; Space Weather; Planets; Comets; Solar Wind; Meteor Showers

**Abstract**

Under Horizon 2020, the Europlanet 2020 Research Infrastructure (EPN2020-RI) will include an entirely new Virtual Access Service, "Planetary Space Weather Services" (PSWS) that will extend the concepts of space weather and space situational awareness to other planets in our Solar System and in particular to spacecraft that voyage through it. PSWS will make five entirely new 'toolkits' accessible to the research community and to industrial partners planning for space missions: a general planetary space weather toolkit, as well as three toolkits dedicated to the following key planetary environments: Mars (in support of the European Space Agency (ESA) ExoMars missions), comets (building on the expected success of the ESA Rosetta mission), and outer planets (in preparation for the ESA JUpiter ICy moon Explorer mission). This will give the European planetary science community new methods, interfaces, functionalities and/or plugins dedicated to planetary space weather in the tools and models available within the partner institutes. It will also create a novel event-diary toolkit aiming at predicting and detecting planetary events like meteor showers and impacts. A variety of tools (in the form of web applications, standalone software, or numerical models in various degrees of implementation) are available for tracing propagation of planetary and/or solar events through the Solar System and modelling the response of the planetary environment (surfaces, atmospheres, ionospheres, and magnetospheres) to those events. But these tools were not originally designed for planetary event prediction and space weather applications. PSWS will provide the additional research and tailoring required to apply them for these purposes. PSWS will be to review, test, improve and adapt methods and tools available within the partner institutes in order to make prototype planetary event and space weather services operational in Europe at the end of the programme. To achieve its objectives PSWS will use a few tools and standards developed for the Astronomy Virtual Observatory (VO). This paper gives an overview of the project together with a few illustrations of prototype services based on VO standards and protocols.

## *1. Introduction*

Planetary Space Weather Services (PSWS) aims at extending the concept of space weather to other planets in our Solar System and in particular to spacecraft that voyage through it. PSWS will give the European planetary scientists for the first time new methods, interfaces, functionalities and/or plug-ins dedicated to planetary space weather in the form of tools and models available in the partner institutes.

Space Weather – the monitoring and prediction of disturbances in our near-space environment and how they are controlled by the Sun - is now recognised as an important aspect of understanding our Earth and protecting vital assets such as orbiting satellites and power grids. The Europlanet 2020 Research Infrastructure (http://www.europlanet-2020-ri.eu/) aims to transform the science of space weather, by extending its scope throughout the Solar System. An entirely new Virtual Access Service, "Planetary Space Weather Services" (PSWS, http://planetaryspaceweather-europlanet.irap.omp.eu/) has therefore been included in the Europlanet H2020 Research Infrastructure funded by the Euroepan Union Framework Programme for Research and Innovation
.
A variety of tools (in the form of web applications, standalone software, or numerical models in various degrees of implementation) are available for tracing propagation of 1) planetary or 2) Solar events through the Solar System and modelling the response of the planetary environment (surfaces, atmospheres, ionospheres, and magnetospheres) to those events. As these tools were usually not designed for 1) planetary event prediction and 2) space weather applications, additional research and tailoring is required to apply them for these purposes. The overall objectives of PSWS will be to review, test, improve and adapt methods and tools available within the partner institutes in order to make prototype 1) planetary event and 2) space weather services operational in Europe at the end of the programme. In particular the aims are:

- To define a service for 1) planetary event and 2) planetary space weather predictions;
- To develop new methods, interfaces, functionalities and/or plug-ins dedicated to planetary space weather in the tools and models already available within the partner institutes;
- To define planetary proxies and reliability factors for planetary space weather applications;
- To validate, compare and enhance the capability of the existing models and tools in order to predict the impact of solar events in the vicinity of Solar System objects;
- To identify user requirements, develop the way to implement event alerts, and chain those to the 1) planetary event and 2) planetary space weather predictions;
- To facilitate discovery or prediction announcements within the PSWS user community in order to watch or warn against specific 1) planetary and 2) planetary space weather events;
- To set up dedicated amateur and/or professional observation campaigns, diffuse contextual information for science data analysis, and enable safety operations of planet-orbiting spacecraft against the risks of impacts from 1) meteors and 2) solar wind disturbances.

The Planetary Space Weather Services will provide 12 services distributed over 4 different service domains – Prediction, Detection, Modelling, Alerts - having each its specific groups of end users. The PSWS portal (http://planetaryspaceweather-europlanet.irap.omp.eu/) gives access to an initial presentation of PSWS activities. Section 2 gives an overview of the foreseen services. Each service will be implemented through a combination of data products, software tools, technical reports, and tutorials. Section 3 describes how the services will comply with Virtual Observatory (VO) methods and standards. Section 4 illustrates some of the VO-compliant functionalities already implemented in some services that are already operational. Section 5 summarizes the status of the project and lists a few perspectives for PSWS services in the VO context and beyond.

## 2. Overview of Planetary Space Weather Services

The Planetary Space Weather Services will provide 12 services distributed over 4 different service domains – Prediction, Detection, Modelling, Alerts. These services are summarized in this section.

### 2.1 Prediction

#### 2.1.1 1D MHD Solar Wind Prediction Tool

The *Centre de Données de Physique des Plasmas* (CDPP) within the *Institut de Recherche en Astrophysique et Planétologie* (IRAP/CNRS) will provide real time and archive access to propagated solar wind parameters at various planetary bodies (Mercury, Venus, Mars, Jupiter, Saturn,…) and spacecraft (Rosetta, Juno, Maven,…) using a 1D magnetohydrodynamic (MHD) code available through the CDPP/AMDA tool (http://amda.cdpp.eu) initially developed by Chihiro Tao (Tao et al., 2005).

#### 2.1.2 Extensions of the CDPP Propagation Tool

The *GFI Informatique* (GFI) will extend the Propagation Tool (Rouillard et al., this issue) available at CDPP (http://propagationtool.cdpp.eu) to the case of comets, giant planet auroral emissions, and catalogues of solar wind disturbances. They will provide new plug-ins including selection of comets as targets, visualization of their trajectories, projection onto solar maps, projection onto J-maps, and estimates of solar wind disturbance arrival times; they will enable the user to use catalogue of solar wind disturbances in order to identify those that have impacted the planetary environments.

#### 2.1.3 Meteor showers

The *Observatoire de Paris* (OBSPARIS) will will link ephemerides of Solar System objects to predictable meteor showers that impact terrestrial planet surfaces or giant planet atmospheres.

#### 2.1.4 Cometary tail crossings

The *Mullard Space Science Laboratory* (MSSL) within the *University College London* (UCL) will develop and post online a software in order to enable users to predict comet tail crossings by any interplanetary spacecraft including future missions like Solar Orbiter.

### 2.2 Detection

#### 2.2.1 Lunar impacts

*Aberystwyth University* (ABER) will upgrade and convert its lunar impact software (https://www.britastro.org/lunar/tlp.htm) and post it online in order to enable users to detect visible flashes in lunar amateur or professional images.

#### 2.2.2 Giant planet fireballs

The *Universidad del Pais Vasco* (UPV/EHU) will upgrade and convert its giant planet fireball detection software (http://pvol2.ehu.eus/psws/jovian_impacts/) and post it online in order to enable users to detect visible fireballs in giant planet amateur or professional images.

#### 2.2.3 Cometary tails

*Mullard Space Science Laboratory* (MSSL) within *University College London* (UCL) will upgrade and convert its comet tail analysis software and post it online, with the aim of also providing it as an interactive suite. The software will be readily accessible to any users (professional or amateur) who work with comet images and wish to obtain an estimate for the solar wind speed at the comet from their observations.

### 2.3 Modelling

#### 2.3.1 Transplanet – Earth, Mars (Venus), Jupiter (Saturn)

The *Centre de Données de Physique des Plasmas* (CDPP) within the *Institut de Recherche en Astrophysique et Planétologie* (IRAP/CNRS) will develop an online version of the hybrid-fluid TRANSPLANET ionospheric model (Marchaudon and Blelly, 2015) that will enable users to make runs on request for Venus, Earth, Mars, Jupiter, and Saturn.

#### 2.3.2 Mars Radiation Environment

*Aberystwyth University* (ABER) together with the Institute of Aerospace Medicine (DLR Cologne) will develop a Mars radiation surface environment model, using modelled average conditions available from Planetocosmics (https://www.spenvis.oma.be/help/models/planetocosmics.html) and synthesised into look-up tables parameterized to variable solar wind conditions at Mars.

#### 2.3.3 Giant planet magnetodiscs

*University College London* (UCL) will adapt the parametric magnetodisc model for Jupiter and Saturn and their space environments in order to take into account realistic, rapid solar wind compressions, based on time-dependent predictions of dynamic pressure from the CDPP Propagation Tool and/or observations of the solar wind at Jupiter orbit.

#### 2.3.4 Jupiter's thermosphere

*University College London* (UCL) will adapt the 2D thermospheric models available for Jupiter and its space environment in order to take into account realistic, rapid solar wind compressions, based on time-dependent predictions of dynamic pressure from the CDPP Propagation Tool and/or observations of the solar wind at Jupiter orbit.

### 2.4 Alerts

The *Observatoire de Paris* (OBSPARIS) together with *University College London* (UCL), the *Institut de Recherche en Astrophysique et Planétologie* (IRAP/CNRS), and the *Space Research Center* (PAS/SRC) will create an Alert service linked to the planetary meteor shower and planetary space weather predictions based on the use of VOEvent (White et al., 2006). This service will be developed in order to facilitate discovery or make predictions within the PSWS user community, in order to watch or warn against specific events. The ultimate objective is to set up dedicated observation campaigns, distribute contextual information for science data analysis, and enable safety operations of planet-orbiting spacecraft against the risks of impacts from meteors or solar wind disturbances.

## 3. Planetary Space Weather Services and the Virtual Observatory

### 3.1 VO protocols of interest for PSWS

We have identified the three VO protocols below for implementation within PSWS services:

- **SAMP** (Taylor et al., 2015), the Simple Application Messaging Protocol is developed now for many years in the frame of the International Virtual Observatory Alliance (IVOA). It enabled easy communication and interoperability between astronomy software, stand-alone and web-based, before being rapidly and largely adopted by the planetary sciences and space physics community. Its attractiveness is based, on one hand, on the use of common file formats for exchange and, on the other hand, on established messaging models.
- **EPN-TAP** (Erard et al., 2014; Erard et al., this issue) is a specific protocol developed by the Europlanet H2020 Research Infrastructure that enables to handle the communication between clients and data providers. Using EPN-TAP, data are accessed in two steps. The first one consists in searching for available EPN-TAP services registered in the IVOA registries, while the second step consists in sending a query searching for data according to specific values of the parameters contained in a table in order to filter the database contents.
- **VOEvent** (Cecconi et al., this issue) is a standardized language used to report observations of astronomical events; it was officially adopted in 2006 by the IVOA. Although most VOEvent messages currently issued are related to supernovae, gravitational microlensing, and gamma-ray bursts, they are intended to be general enough to describe all types of observations of astronomical events.

### 3.2 Implementation of VO protocols within PSWS

The three VO protocols identified previously will be implemented in the PSWS services as described in the Table below.

| PSWS Services | Type of developments | Use of EPN-TAP | Use of SAMP | Use of VOEvent |
|---|---|---|---|---|
| 1D MHD Solar Wind Prediction Tool | Website+Database+Alerts | yes | yes | yes |
| Propagation Tool | Software+Database+Alerts | yes | yes | yes |

| Meteor showers | Website+Alerts | no | no | yes |
|---|---|---|---|---|
| Cometary tail crossings | Software+Alerts | no | no | yes |
| Lunar impacts | Software+Database+Alerts | yes | yes | yes |
| Giant planet fireballs | Software+Database+Alerts | yes | yes | yes |
| Cometary tails | Software+Database | yes | yes | no |
| Transplanet | Runs on request+Database | yes | yes | no |
| Mars Radiation Environment | Runs on request+Database | yes | yes | no |
| Giant planet magnetodiscs | Runs on request+Database | yes | yes | no |
| Jupiter's thermosphere/ionosphere | Runs on request+Database | yes | yes | no |
| Alerts | Database+Alerts | yes | no | yes |

Table 1. PSWS services, type of developments, and VO protocols to be implemented within the corresponding services.

## 4. Planetary Space Weather VO Services Illustrated

### 4.1 1D MHD Solar Wind Prediction Tool

The lack of solar wind monitoring just upstream of Solar System bodies can be overcome by using simulations that propagate the solar wind from in situ observations obtained elsewhere in the Solar System (e.g., Zieger and Hansen, 2008). We have implemented in the CDPP tools AMDA (Automated Multi-Dataset Analysis, Génot et al., 2010), 3DView (Génot et al., this issue), and Propagation Tool (Rouillard et al., this issue), a one-dimensional (1D) magnetohydrodynamic (MHD) simulation of the solar wind propagation developed by Tao et al. (2005). The simulation uses time-varying boundary conditions at one Astronomical Unit (AU) obtained either from hourly solar wind data observed near the Earth made available by OMNI (http://omniweb.gsfc.nasa.gov/), or real-time ACE observations. Simulated propagated solar wind parameters such as density, temperature, velocity, dynamic pressure, and tangential magnetic field in RTN coordinates are available for various planetary bodies (Mercury, Venus, Earth, Mars, Jupiter, Saturn, and comet Churyumov-Gerasimenko) and spacecraft (e.g., Rosetta, Juno). The use of real-time ACE observations will enable us in the near future to make predictions for the potential interactions of solar wind disturbances with planetary bodies.

Figure 1 displays in the AMDA tool observed and simulated dynamic pressure at Earth, Jupiter, and Saturn for a time period when the three planets are aligned in November/December 2000 (Prangé et al., 2000). We can see at the beginning of the time interval a series of CMEs observed in situ at Earth that propagate in the Solar System, evolve, and eventually merge before to reach Jupiter in the middle of the time interval as well as Saturn at the end of the time interval as predicted by the 1D MHD simulation.

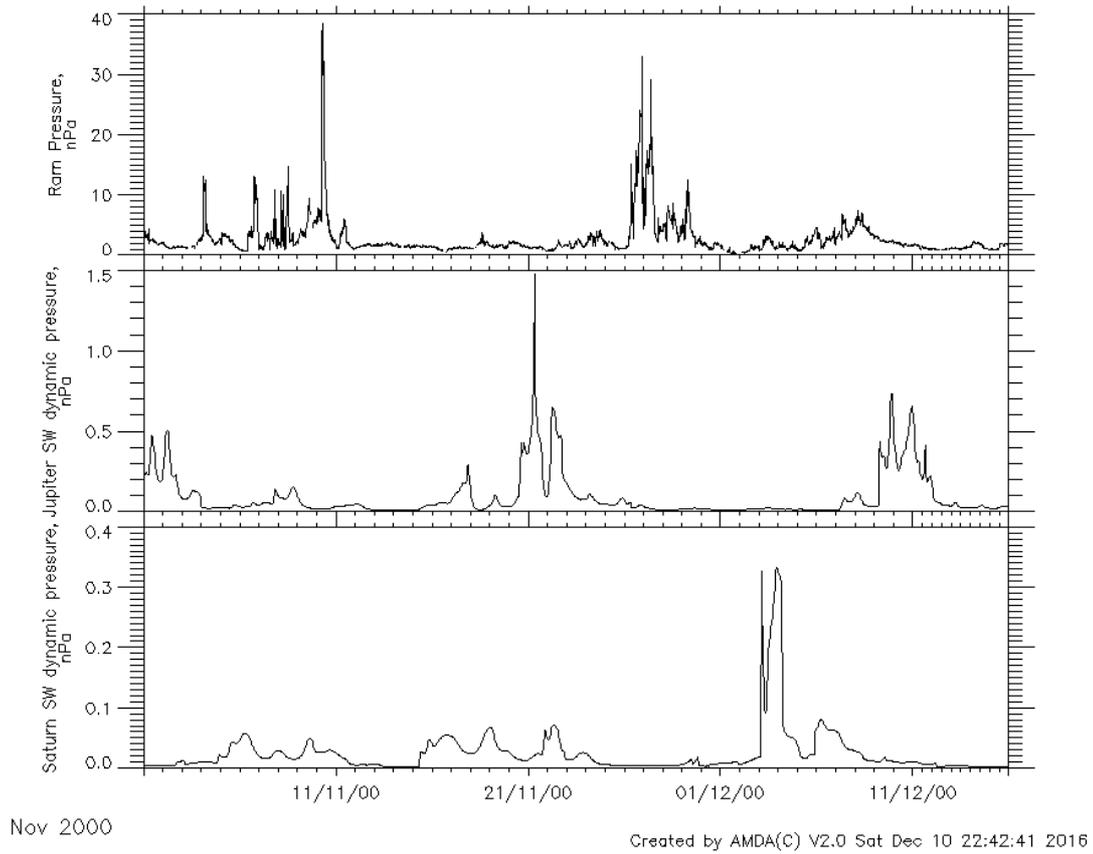

Figure 1. Dynamic Pressure (in nPa) versus time: OMNI observations at Earth (top), 1D MHD simulations at Jupiter (middle), and Saturn (bottom). Plot done with the AMDA tool.

Figure 2 illustrates how the EPN-TAP protocol can be used from the VESPA interface (Erard et al., this issue) in order to search for remote Hubble Space Telescope observations of giant planets ultraviolet aurorae available in APIS (Lamy et al., 2015) during the same time period.

Figure 2. Results of an EPN-TAP query from the VESPA interface searching for Saturn observations during the time period 01/11/2000-15/12/2000 in the APIS service.

In order to facilitate similar queries an APIS interface has been directly implemented within the AMDA and Propagation Tool (Figure 3). These two tools are connected through the SAMP Protocol with more sophisticated astrophysical VO tools such as Aladin in order to enable further visualization and analysis of the combined remote and in situ observations.

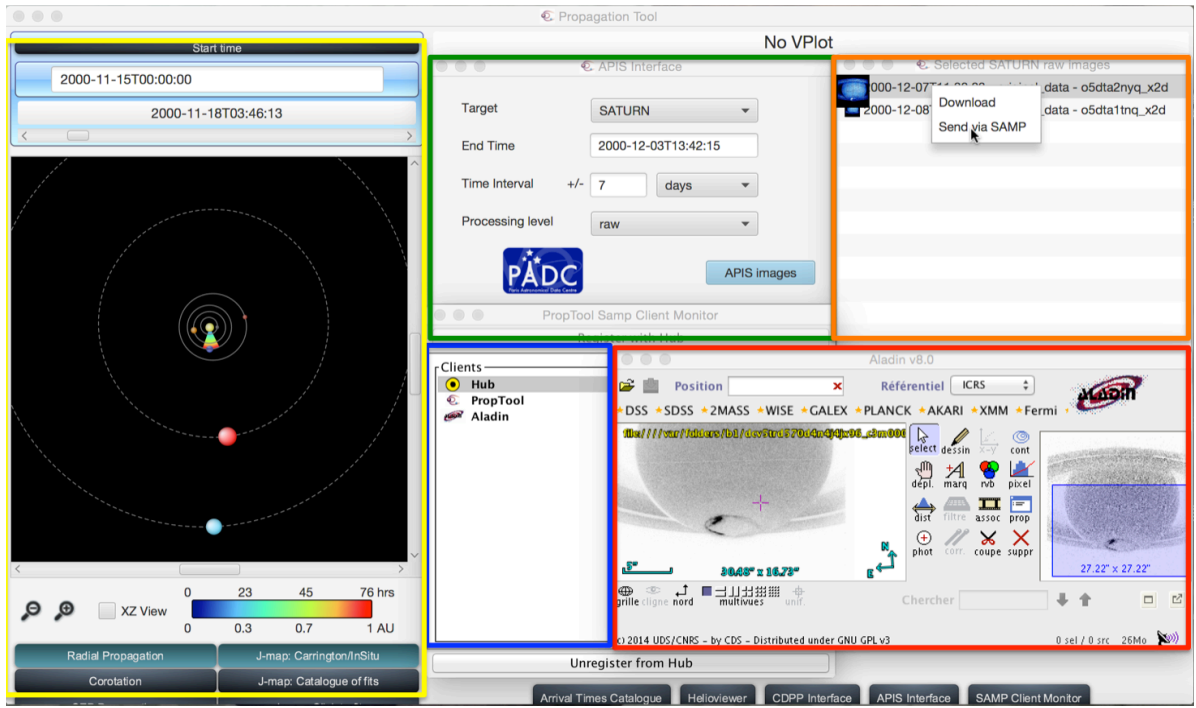

Figure 3. Illustration of some VO functionalities available in the CDPP Propagation Tool. Yellow box: Location of the planets in the heliosphere (Earth in dark blue, Jupiter in red, Saturn in light blue) for the time period 15/11/2000-18/11/2000. Green box: APIS interface to query Hubble Space Telescope giant planet auroral observations using the EPN-TAP protocol. Orange box: results of the query. Blue box: SAMP client monitor activated. Red box: visualization in the Aladin tool of a giant planet auroral observation from the APIS database enabled by SAMP.

### 4.2 Extensions of the CDPP Propagation Tool to comets

The structure and dynamics of cometary plasma tails witness the solar wind variability. Cometary observations are particularly attractive for amateur astronomers who usually are the first witnesses of remarkable solar wind – comet interactions. We have extended the CDPP 3DView (http://3dview.cdpp.eu/, Génot et al., this issue) and Propagation Tool in order to include comets as targets so that for example the Propagation Tool can be used in order to estimate the properties of the solar wind as well as identify the arrival of a solar wind disturbance in the vicinity of a particular comet.

On 20 April 2007, a tail disconnection event on comet 2P/Encke caused by a coronal mass ejection (CME) was observed by the STEREO-A spacecraft (Vourlidas et al., 2007). Figure 4 illustrates how the CDPP Propagation Tool can be used in order to identify on a J-map the interaction of the comet with the CME. The corresponding J-map is generated by extracting bands of pixels in STEREO-A coronal and heliospheric images along the ecliptic planes and stacking them vertically (along the ordinate) with time (along the abscissae).

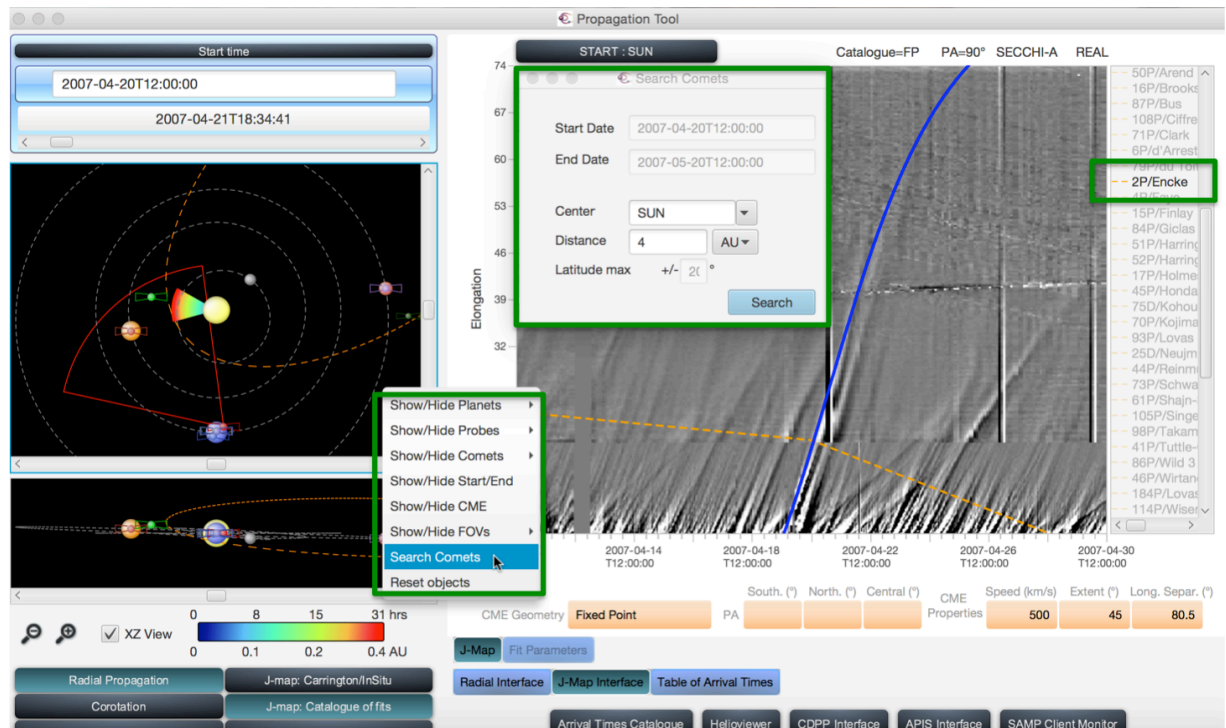

Figure 4. Illustration of the new PSWS functionality implemented within the CDPP Propagation Tool and dedicated to comets. Green box: search window to identify comets flying by a central body at a particular distance during the selected time period. Once a comet is identified and selected (here 2P/Encke), its orbit is displayed on the left. Its location in a J-map obtained from STEREO-A observations is displayed on the right.

### *4.3 Extensions of the Propagation Tool to catalogues of solar wind disturbances*

The advent of wide-angle imaging of the inner heliosphere has revolutionised the study of the solar wind and, in particular, transient solar wind structures such as Coronal Mass Ejections (CMEs) and Co-rotating Interaction Regions (CIRs). With Heliospheric Imaging came the unique ability to track the evolution of these features as they propagate throughout the heliosphere. The FP7 project HELCATS (HELiospheric Cataloguing, Analysis, and Technique Service, https://www.helcats-fp7.eu/) aims to producing a definitive catalogue of CMEs imaged by the Heliospheric Imager (HI) instruments onboard the NASA STEREO spacecraft. Outputs of HELCATS Work Packages 3 (CME KINEMATICS Catalogue) and 5 (STEREO SIR/CIR Catalogue) have been included in the CDPP Propagation Tool. This functionality is illustrated in Figure 5 and will be extended to additional catalogues publicly available when possible.

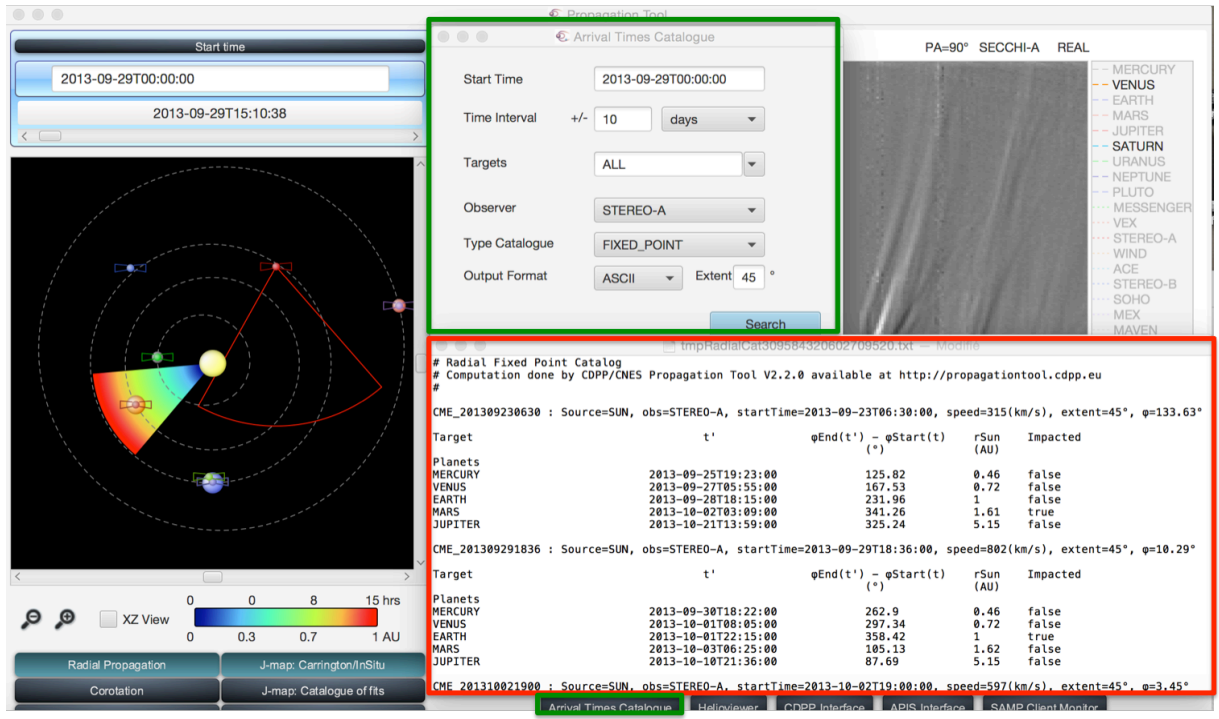

Figure 5. Illustration of the new PSWS functionality implemented within the Propagation Tool and dedicated to the ingestion of catalogues of solar wind disturbances. These catalogues can be used to predict arrival times of each disturbance at various Solar System bodies and spacecraft. Green box: interface dedicated to the choice of catalogues based on STEREO-A, STEREO-B, and SOHO observations. Red box: output of the query showing which planet is impacted (true) by a particular Coronal Mass Ejection (CME) from the chosen catalogue during the user-defined time period.

### *4.4 Giant planet fireballs*

Meteors impacting Jupiter's upper atmosphere can create spectacular fireballs. Relatively small objects left over from the formation of the solar system 4.5 billion years ago still hit Jupiter frequently. The resulting impacts are bright enough so that amateur astronomers can serendipitously detect them. Four fireballs were first reported by amateur astronomers in July 2009, in 2010 (on June 3[rd] and on August 20[th]) and in 2016 (on March 17[th], Figure 6) before to be detailed by professional astronomers following on the initial amateur observations (e.g., Hueso et al., 2010). Groups of amateurs worldwide have then coordinated efforts in order to obtain improved estimates of the number of small bodies around Jupiter and how they interact with the planet. Dramatic impacts with Jupiter can indeed be captured with standard amateur equipment and analysed with easy-to-use software. Within PSWS, such software (http://pvol2.ehu.eus/psws/jovian_impacts/) will be further enhanced in order to improve their usability and reach an even wider participation by amateurs. It will be connected to the Planetary Virtual Observatory and Laboratory (PVOL, Hueso et al., this issue) that provides a searchable database of ground-based amateur observations of solar system planets (http://pvol2.ehu.eus/pvol2/).

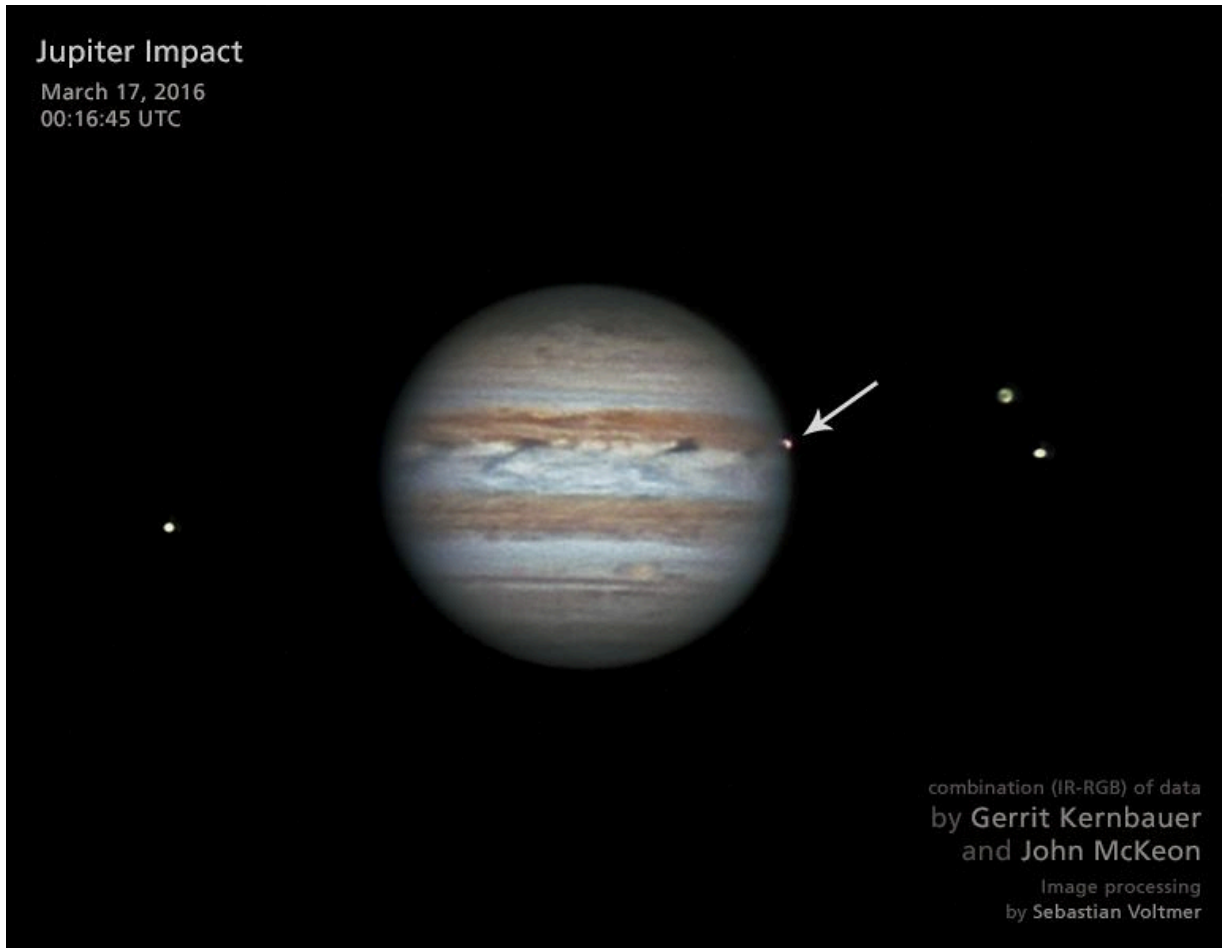

Figure 6. 2016 March 17th fireball captured by Gerrit Kernbauer and John McKeon. Image processed by Sebastian Voltmer. Credit: G. Kernbauer, J. McKeon, S. Voltmer.

## 5. Conclusions

Planetary Space Weather Services (PSWS) within the Europlanet H2020 Research Infrastructure are currently developed following protocols and standards available in Astrophysical, Solar Physics and Planetary Science Virtual Observatories (VO). Several VO-compliant functionalities already implemented in various tools as well as in development have been described in the present paper. We aimed at showing how planetary space weather can benefit from the VO. The proposed Planetary Space Weather Services will be accessible to the research community, amateur astronomers as well as to industrial partners planning for space missions dedicated in particular to the following key planetary environments: Mars, in support of ESA's ExoMars missions; comets, building on the success of the ESA Rosetta mission; and outer planets, in preparation for the ESA JUpiter ICy moon Explorer (JUICE). These services will also be augmented by the future Solar Orbiter and BepiColombo observations. This new facility will not only have an impact on planetary space missions but will also allow the hardness of spacecraft and their components to be evaluated under variety of known conditions, particularly radiation conditions, extending their knownflight-worthiness for terrestrial applications.

In addition to their connections with the Virtual Observatory, the Planetary Space Weather Services developed within the Europlanet H2020 Research Infrastructure will be strongly linked to the space weather services developed within the ESA's Space Situational Awareness programme (http://swe.ssa.esa.int/heliospheric-weather). Connections with the CCMC (Community Coordinated Modeling Center, http://ccmc.gsfc.nasa.gov/) in the United States will also be studied in the future.

## Acknowlegments

Europlanet 2020 RI has received funding from the European Union's Horizon 2020 research and innovation programme under grant agreement No 654208.

## References

Cecconi, B. et al., Developing an Efficient Planetary Space Weather Alert Service using Virtual Observatory Standards, this issue

Erard, S. et al., The EPN-TAP protocol for the Planetary Science Virtual Observatory, Astronomy and Computing, 7, 2014

Erard, S. et al., The EPN-TAP protocol for the Planetary Science Virtual Observatory, Astronomy and Computing, 7, 52-61, 10.1016/j.ascom.2014.07.008, 2014

Erard, S. et al., VESPA: a community-driven Virtual Observatory in Planetary Science, this issue

Génot, V. et al., Joining the yellow hub: Uses of the Simple Application Messaging Protocol in Space Physics analysis tools, Astronomy and Computing, 7, 2014

Génot, V. et al., Génot, V., Jacquey, C., Bouchemit, M., Gangloff, M., Fedorov, A. Lavraud, B., André, N., Broussillou, L., Harvey, C., Pallier, E., Penou, E., Budnik, E., Hitier, R., Cecconi, B., Dériot, F., Heulet, D., Pinçon, J.-L., Space Weather applications with CDPP/AMDA, Advances in Space Research, 45, 9, 1145-1155, 2010

Hueso, R. et al., First Earth-based Detection of a Superbolide on Jupiter, Ap J Letters, 721, L129, 2010

Hueso, R. et al., The Planetary Virtual Observatory and Laboratory, this issue

Lamy, L., Prangé, R., Henry, F., Le Sidaner, P., The Auroral Planetary Imaging and Spectroscopy (APIS) service, Astronomy and Computing, Volume 11, 138-145, 10.1016/j.ascom.2015.01.005, 2015

Lilensten, J. et al., What characterizes planetary space weather?, The Astronomy and Astrophysics Review, 22, 79, 10.1007/s00159-014-0079-6, 2014

Marchaudon, A., and P.-L. Blelly, A new interhemispheric 16-moment model of the plasmasphere-ionosphere system: IPIM, Journal of Geophysical Research: Space Physics, 120, 5728-5745, 10.1002/2015JA021193, 2015

Plainaki, C. et al., Planetary space weather: scientific aspects and future perspectives, Journal of Space Weather and Space Climate, 6, A31, 10.1051/swsc/2016024, 2016

Prangé, R., L. Pallier, K.C. Hansen, R. Howard, A. Vourlidas, R. Courtin, C. Parkinson, An interplanetary shock traced by planetary auroral storms from the Sun to Saturn, Nature, 432,

7013, 10.1038/nature02986, 2004

Rouillard, A. et al., A propagation tool to connect remote-sensing observations with in-situ measurements of heliospheric structures, this issue

Tao, C. et al., Magnetic field variations in the Jovian magnetotail induced by solar wind dynamic, Journal of Geophysical Research: Space Physics, 110, A11208, 10.1029/2004JA010959, 2005

Taylor, M.B., et al., SAMP, the Simple Application Messaging Protocol: Letting applications talk to each other, Astronomy and Computing, 11, 81-90, 2015

Vourlidas, A., et al., First Direct Observation of the Interaction between a Comet and a Coronal Mass Ejection Leading to a Complete Plasma Tail Disconnection, The Astrophysical Journal, 668, L79-L82, 10.1086/522587, 2007

White, R.R., et al., Astronomical network event and observation notification, Astronomische Nachrichten, 327, 8, 775, 10.1002/asna.200610631, 2006

Zieger, B., and K.C. Hansen, Statistical validation of a solar wind propagation model from 1 to 10 AU, Journal of Geophysical Research: Space Physics, 113, A08107, 10.1029/2008JA013046, 2008